\documentstyle[epsf,nato]{crckapb}

\newcommand{\be}{\begin{equation}}
\newcommand{\ee}{\end{equation}}
\newcommand{\bea}{\begin{eqnarray}}
\newcommand{\eea}{\end{eqnarray}}
\newcommand{\no}{\noindent}

\newcommand{\Tr}{{\rm Tr}}
\def\NPB{{\it Nucl.\ Phys.}{\bf  B}}
\def\NPBPS{{\it Nucl.\ Phys.}\ [Proc.\ Suppl.]}
\def\PLB{{\it Phys.\ Lett.}{\bf  B}}
\def\PRL{{\it Phys.\ Rev.\ Lett.}}
\def\CMP{{\it Comm.\ Math.\ Phys.}}

\def\RMP{{\it Rev.\ Mod.\ Phys.}}
\def\PRD{{\it Phys.\ Rev.}{\bf  D}}

\begin{opening}
\title{ON THE TOPOLOGICAL STRUCTURE OF THE QCD VACUUM}
\subtitle{Problems and Results of Lattice Studies}

\author{Ion-Olimpiu Stamatescu}
\institute{ITP, U. Heidelberg,
           Philosophenweg 16, D-69120 Heidelberg}
\institute{FESt,
           Schmeilweg 5, D-69118 Heidelberg, Germany}

\end{opening}

\runningtitle{TOPOLOGICAL STRUCTURE OF THE QCD VACUUM}

\begin{document}

\no {\bf Abstract}: A review of results from lattice studies using improved and scale
controlled cooling methods is presented and their significance is
discussed. The improvement of the action ensures stable instanton
solutions of physical sizes. The scale controlled cooling can be
generally used as a gauge invariant low pass filter to extract the
physics from noisy MC configurations; in particular it preserves
instanton-antiinstanton pairs selected according to their interaction.
We apply these methods to analyze various features of the topological
structure of the Yang-Mills vacuum in a scaling invariant way.


\section{Introduction}

This article presents results from a series of works involving 
besides the above author the following persons in various
 combinations:

\no PH.~DE~FORCRAND (ETH Z\"urich and TH-Div., CERN, Switzerland),

\no M.~GARC\'IA~P\'EREZ (Dept.  F\'{\i}s. Teor., Univ. Aut. de Madrid, Spain),

\no T.~HASHIMOTO (Dept. of Appl. Physics, Fukui Univ.,  Japan),

\no J.E.~HETRICK (Physics Dept., U. of the Pacific, Stockton, USA),

\no S.~HIOKI (Dept. of Physics, Tezukayama Univ., Nara, Japan),

\no E.~LAERMANN (FB Physik, Univ. Bielefeld, Germany),

\no J.F.~LAGAE (HEP Div., Argonne Nat. Lab., USA),

\no H.~MATSUFURU (RCNP, Osaka, Japan),

\no O.~MIYAMURA, T.~UMEDA (Dept. of Physics, Hiroshima Univ., Japan),

\no A.~NAKAMURA (RIISE, Hiroshima Univ.,  Japan),

\no O.~PHILIPSEN (Th. Div., CERN, G\'en\`eve, Switzerland),

\no T.~TAKAISHI (Hiroshima University of Economics, Japan).

\no The simulations and results on which the following discussions are based 
are described in 
\cite{mnpnp},\cite{latt97},\cite{buck},\cite{latt98},\cite{ric},\cite{latt99},\cite{tren}.

\section{Physical Questions and Problems of the Analysis}

Instantons are self-dual solutions of the
Euclidean field equations  mapping the $SU(2)$ color group space
on the 3-d sphere at spatial infinity (for  $SU(N_c)$ 
the $SU(2)$ subgroups are relevant). They are defined among other by
 winding number (topological charge) and  position and size parameters.
The ${\bf R}^4$ (anti-)instanton (A)I has 
charge (q) and action (s) densities \cite{bela}:

\bea
 q(x) = \frac{1}{8 \pi^2}F^*F =
{6 Q\over { \pi^2\rho^4}} \left[ 1 + 
\sum_{\mu=1}^4\left( {{x_{\mu}-x_{\mu}^0} \over {\rho}}\right)^2
\right]^{-4},\ s(x) = S_0 |q(x)|, 
\label{e.dens}\\
Q=\int d^4x q(x) = \pm {\rm integer},\ \ 
S=\int d^4x s(x) = |Q|S_0,\ \ S_0=8 \pi^2.\nonumber
\eea
In ${\bf T}^4$  only $|Q|\ne 1$ exist  \cite{brvbaal}. 
In ${\bf R}^3{\bf T}$ 
(relevant at non-zero temperature $T$) the ``caloron"
 \cite{calor} describes objects flattening 
in the Euclidean time direction and of limited spatial size
$\rho \le \sqrt{3}/\pi T$. 
Finally, various kinds of twisted boundary conditions allow for new
 solutions \cite{MP},\cite{tw2}.
Superpositions of $N$ 
I's {\it or}  A's also 
correspond to (higher) minima of the action: $S = N S_0$.   
A pair I-A is not, however, a minimum  and 
  $S^ {IA}$ depends also on ``overlap" $\omega$:

\be
S^{\rm IA}=2S_0 -S_{\rm int}^{\rm IA} < 2S_0, \ \ 
\omega = (\rho_{\rm I}+ \rho_{\rm A})/ d_{\rm IA},\ 
d_{\rm IA} = |x_{\rm I}^0 - x_{\rm A}^0|. 
\label{e.over}
\ee

\no  Effects associated with instantons are (for a review see
\cite{Scha}):\par
\no -  the $U_{\rm A}(1)$ symmetry breaking (via 
the Witten - Veneziano formula\cite{wive}), \par
\no - chiral symmetry breaking (via zero modes of the Dirac operator),\par
\no - dynamical effects for the physics at intermediate distances.

\no The first one  only involves  global topological 
properties (susceptibility).
The other possible effects depend on details of the local structure, like
density and size distribution of  instantons, and the
calculation  involves models. 
In the {\it instanton liquid model} \cite{Shury}, 
for instance, which is the standard model of instanton physics,
various predictions  have been made \cite{Scha}.
Numerical simulations
should test the ingredients used in these models.
Thereby:

\no - The instantons can be in a gas,
 liquid, or crystalline phase. 
 The diluteness is expressed by 
the ``packing fraction" 
\be
f=\frac{\pi^2 \langle\rho^4\rangle N}{2V}.
\label{e.pac}
\ee
\no - The size distribution controls diluteness and I-R properties.
There is no prediction for  large sizes. For small sizes 
the dilute gas approximation  
gives:
\be
P(\rho) \sim  \rho^p, \ \ p_{\rm dilute} = -5 + b,\ \ b = 11 N_c/3
\label{e.rd1}
\ee
\no - Denoting by $N_{\rm I}  (N_{\rm A})$ the number of I's (A's)
in a configuration, and with $N = N_{\rm I} + N_{\rm A}$, $Q=N_{\rm I}-N_{\rm A}$
we  write:

\be
c_{\rm P} \equiv \left({\langle N^2 \rangle - \langle N \rangle^2}
\right) /{\langle N \rangle}.
 \label{e.corr}
\ee

\no 
For a poissonian distribution $c_{\rm P} = 1$, while low
energy sum rules suggest $c_{\rm P} = {4 \over b}$ \cite{IMP}. 
A further measure for the
character of the I,A distribution is:

\be
c_{\rm int}=\frac{\langle N_{\rm I}N_{\rm A}\rangle-\langle N_{\rm I}\rangle \langle N_{\rm A}\rangle}
{\sqrt{(\langle N_{\rm I}^2\rangle-\langle N_{\rm I}\rangle^2)
(\langle N_{\rm A}^2\rangle-\langle N_{\rm A}\rangle^2)}} =
\frac{\langle N^2\rangle-\langle Q^2\rangle - \langle N\rangle^2}
{\langle N^2\rangle+\langle Q^2\rangle - \langle N\rangle^2}
\label{e.cint}
\ee

\no which tests the correlation between I's and A's: $c_{\rm int}$ is 0 if
there is no such correlation, but goes to 1 if I and A appear correlated
(e.g., pairwise).

For non-perturbative analysis one resorts to lattice simulations.
The Monte Carlo method is sometimes considered
to be a tool which produces numerical results, 
without improving our understanding  of the physical
mechanisms.  This view, however, has turned out to be
too narrow and different structural questions 
have been investigated by means of Monte Carlo calculations.
The problems such an analysis of the topological structure
encounters are twofold. {\it Problems of principle} are:

- UV-divergences in correlation functions like
$\langle q(x)q(0)\rangle$ which contain
 the susceptibility information 
in a contact term
(notice that
$\langle q(x)q(0)\rangle < 0$ for 
disjoint supports, e.g. for $x \geq 2a$ if 
$F_{\mu \nu}$ is defined on plaquettes) \cite{erha}. 

- Dislocations: concentration
of topological charge on one or few plaquettes. They have action
smaller than $S_0$ and  may be produced copiously, spoiling the measurement
of $Q$ and the continuum limit \cite{pute}.

- Since $S_{\rm int}^{\rm IA}< 2S_0$   close pairs  can be
easily produced. With increasing overlap it becomes difficult to define genuine 
I-A pairs as distinguished from short range density fluctuations and the description
 in terms of I's and A's becomes questionable \cite{buck}.

It appears therefore that a certain amount of smoothing may be necessary 
to define topology on the lattice. {\it Practical problems} are:

- The need to develop smoothing methods which can be controlled in a
physical way. They should be based on independent physical arguments,
leave the investigated structure undistorted and 
allow to observe its properties without
making  assumptions about them. In particular
 scaling behaviour of topological distributions 
should not be assumed but revealed.

- The description of the topological structure in terms of
I's and A's. Overlapping structures, especially with opposite charges, become
difficult to fit with a superposition ansatz based on (\ref{e.dens}). 
One needs to introduce various criteria and 
some measure for the adequacy of the description which, however, remains imperfect.

\section{Scale Controlled Cooling [5],[6]}  

The problem of developing good smoothing methods has been approached in 
a number of ways: standard cooling, smearing, blocking, combination of them 
and others \cite{tep0},\cite{mpg},\cite{boul0},\cite{boul}.
The smoothing is controlled by monitoring, by giving a
blocking step or by tuning smearing to approximate a RG transformation
with fixed point action etc. The most natural control for a smoothing method,
however, is by giving the smoothing radius or wave length threshold and we should like, therefore, to develop 
a method systematically controlled by a physical scale. 
We  start from cooling, 
since this procedure is analytically
defined (by reference to the equations of motion) and has as fixed point 
nontrivial semi-classically relevant configurations with $N=|Q|$.
Cooling, which proceeds by local minimization of a given lattice action, 
works as a diffusion process, smoothing out
 increasingly large regions
 \cite{tep1}. Thereby properties of the physical spectrum measured from
 correlations at a given distance are affected, 
 in particular the string tension drops
rapidly with the number of cooling sweeps $n_c$ \cite{dig89}. Instantons
 may be distorted and lost during cooling by bad scaling properties of the
action and also by I-A annihilation. The  
susceptibility may be spoiled by dislocations. We thus need:\par
\no 1) To use an action with
 practically scale invariant instanton solutions and a sharp dislocation threshold
{\it fixed in lattice units} to eliminate UV noise (dislocations) while
shrinking in approaching continuum, and\par
\no 2) To ensure that cooling is controlled by
 a physical smoothing scale such that the structure at larger scales
remains unchanged. Monitoring and
engineering should not be necessary, since this may
introduce arbitrariness. 

The  {\it Restricted Improved Cooling} (RIC) introduced in  \cite{ric}
fulfills these requirements. RIC acts as a {\it gauge invariant low pass filter}
preserving physics at scales above a 
{\it cooling radius} $r_c$ which can be fixed unequivocally,
while smoothing out the structure below $r_c$.
In particular the string tension is preserved beyond $r_c$,
instantons are stable, dislocations are eliminated, and I-A
 pairs are retained above an overlap threshold 
depending on $r_c$.

\begin{figure}[tb]
\vspace{11.5cm}
\includegraphics{f1a.ps}
\includegraphics{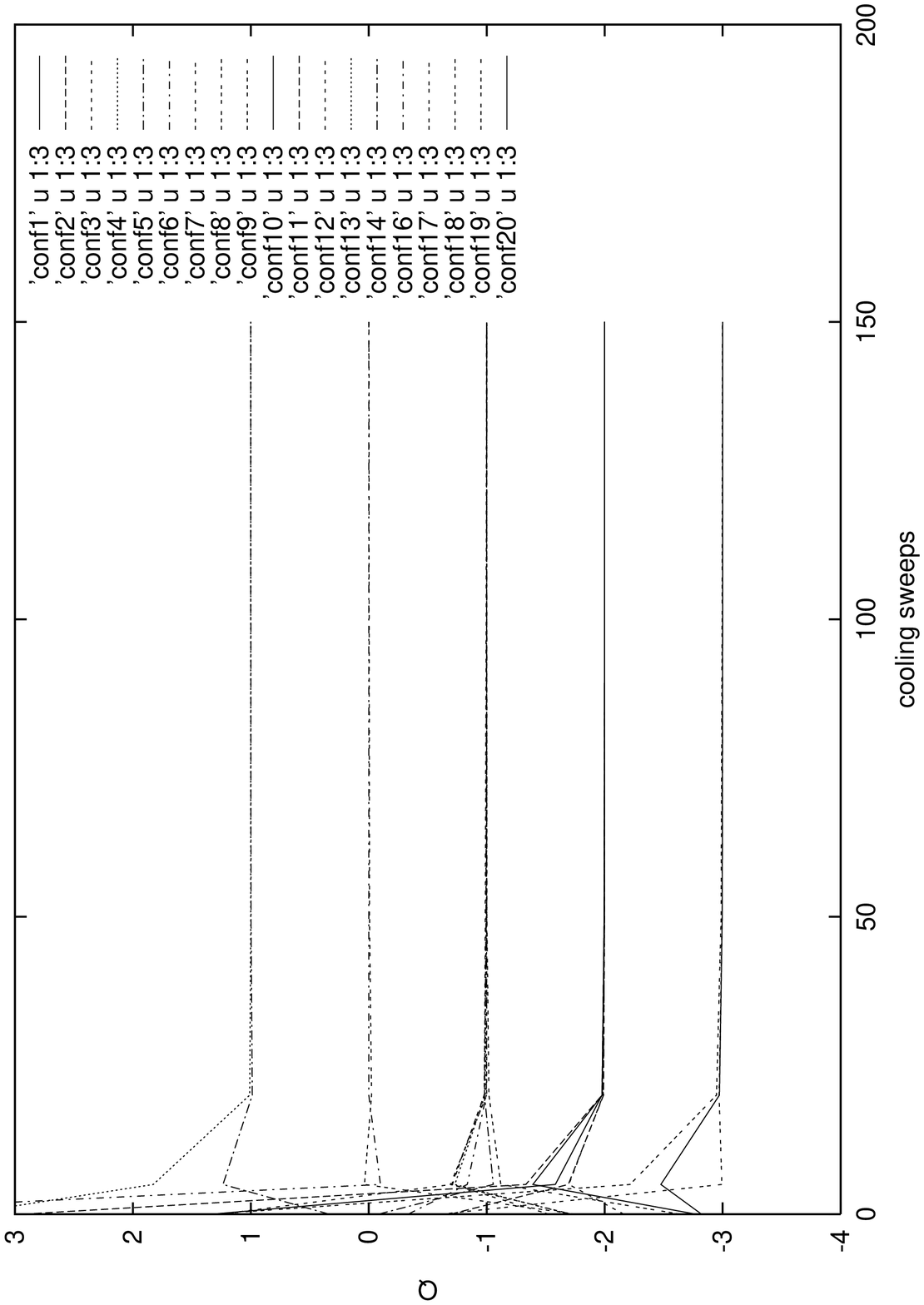}
\includegraphics{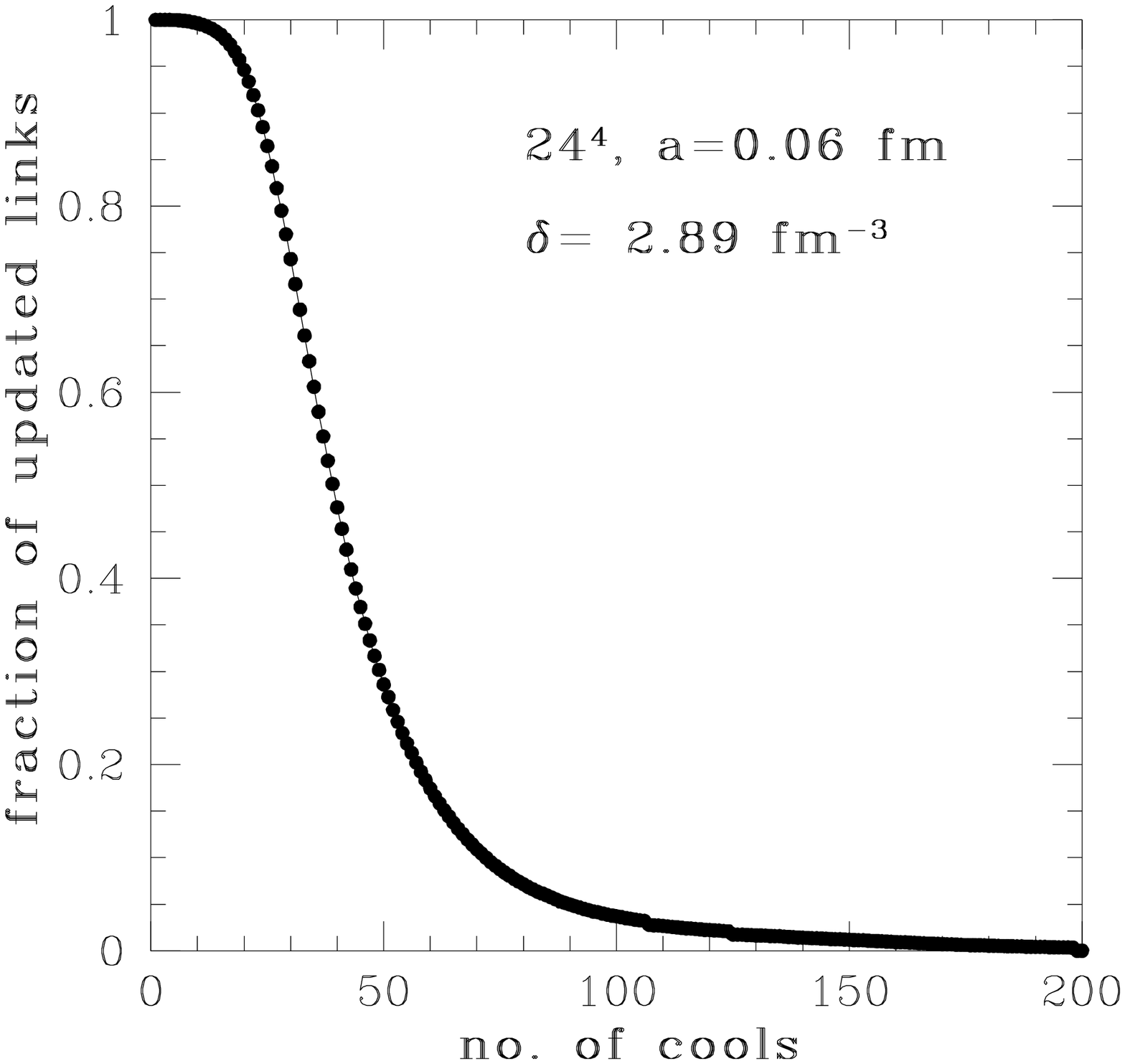}
\caption{IC and RIC properties: Evolution of sizes of single instantons
(upper left).
Improved (5Li) and Wilson action {\it vs} instanton size 
given in lattice units; $\rho_{peak}$ is defined in  [1] (upper right).
Evolution of the improved charge $Q$ for 20 $SU(3)$ MC configurations [7]
(lower left). Saturation of cooling for RIC: fraction of updated links as
function of the cooling sweeps -- for stopped standard cooling or IC this 
would be a step function (lower right).}
\vspace{-0.4cm}
\label{f.ic}
\end{figure}

RIC uses the action of the {\it Improved Cooling} algorithm (IC)
with 5 planar, fundamental Wilson loops \cite{ic},\cite{mnpnp}.
This action is 
 correct to order ${\cal O}(a^6)$ 
and has a {\it dislocation threshold} 
$\rho_0 /a ={\hat \rho}_0 \sim 2.3$, below which short range topological structure 
is smoothed out (note that $\rho_0 \rightarrow 0$ in approaching continuum).
Above  $\rho_0$, 
instantons are stable under IC to
any degree of cooling (however, I-A pairs annihilate).  The corresponding
{\it improved charge density} using the same combination of loops
leads to an integer charge already after a few cooling
sweeps and  stable thereafter. See Fig. \ref{f.ic}.

 Recall that the cooling  
algorithm is derived from the equations of motion
\be
\label{eq}
U_\mu(x) W_\mu(x)^\dagger - W_\mu(x) U_\mu(x)^\dagger = 0  ,
\ee
\no where $W$ is the sum of staples connected to the link $U_\mu(x)$ 
in the action, as (we  restrict here to $SU(2)$, for the general case 
see \cite{ric}):
\be
U \rightarrow U' = V = {W}/{||W||} , \ ||W||^2=\frac{1}{2} \Tr(W W^\dagger).
\label{e.sta}
\ee
\no We define RIC by the  constraint
that only those links be updated, which violate the equation of motion by
more than some
chosen threshold\footnote{We thank 
F. Niedermayer for this suggestion.}:
\be
U \rightarrow V \ \  {\rm iff} \ \  
\Delta_\mu(x)^2= {a^{-6}} {\rm Tr}(1-U V^{\dagger})\geq\delta^2 .
\label{cond}
\ee
\no We have
$\Delta_\mu^2 (x) \propto 
-\Tr((D_\nu F_{\nu \mu}(x))^2)$ 
 in  continuum limit \cite{map}. 
 Thus 
$\delta$ controls the energy 
of the fluctuations around classical solutions 
and acts as a filter for short wavelengths. 
Since it uses the same action RIC has the same scaling properties as IC. 
However, since RIC  does not update links  already close to a solution,
it changes fewer links after
every iteration until  the algorithm saturates -- see Fig. \ref{f.ic}.
(\ref{cond}) defines a constrained minimization and 
the smoothing is 
 homogeneous over the lattice (see \cite{ric}).

Since the parameter $\delta$ which defines the 
cooling is already a physical quantity it should be related 
to a physical scale. For Yang Mills theory the latter 
 involves the string tension $\sigma$. We calculate the ``effective mass"
 $M(t)$ from  
correlation functions of spatial
Polyakov loops separated by $t$ steps in time. Asymptotically 
$M(t) \simeq N_s\sigma$ up to finite size corrections.

\begin{figure}[htb]
\vspace{5cm}
\includegraphics{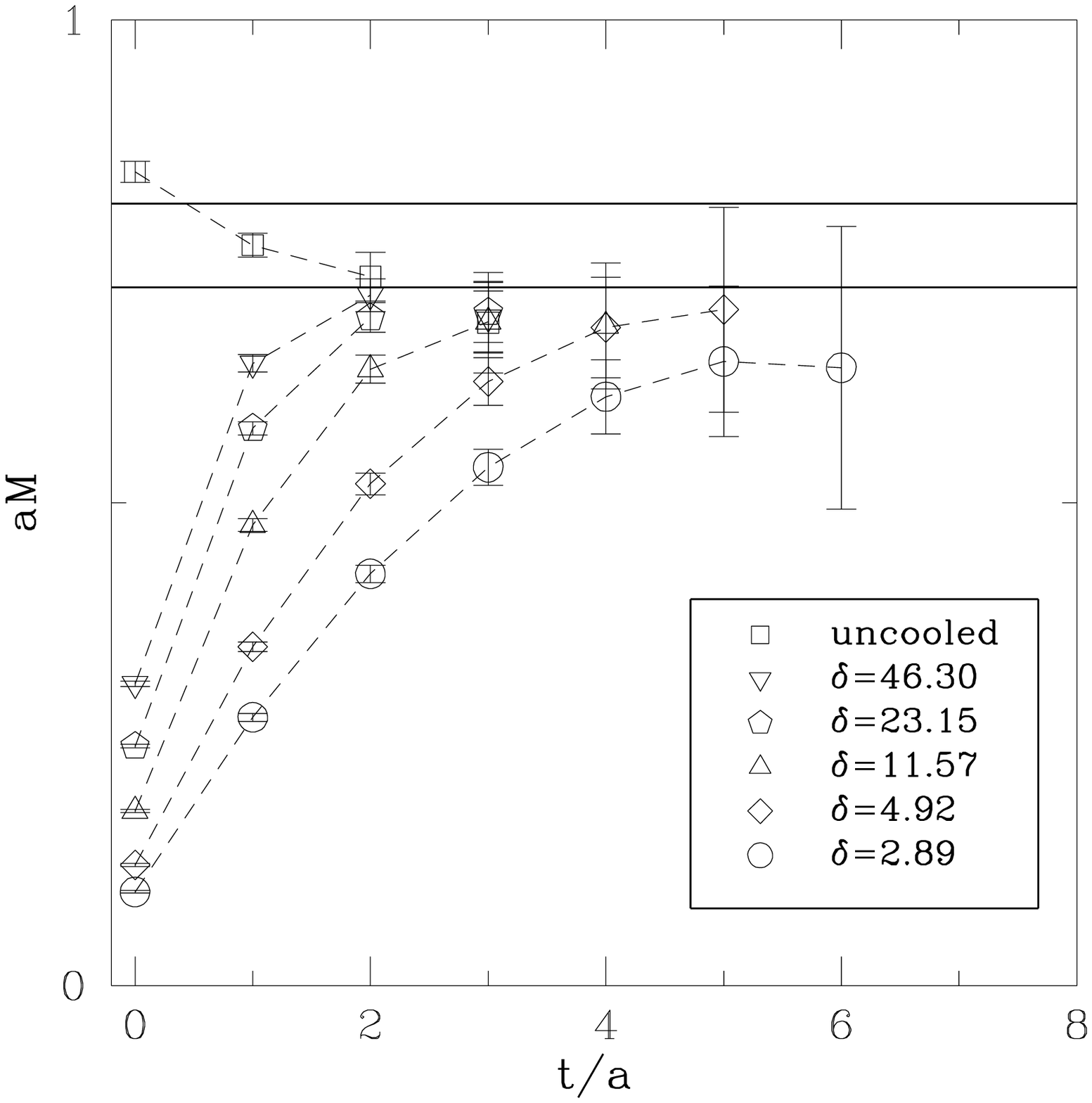} 
\includegraphics{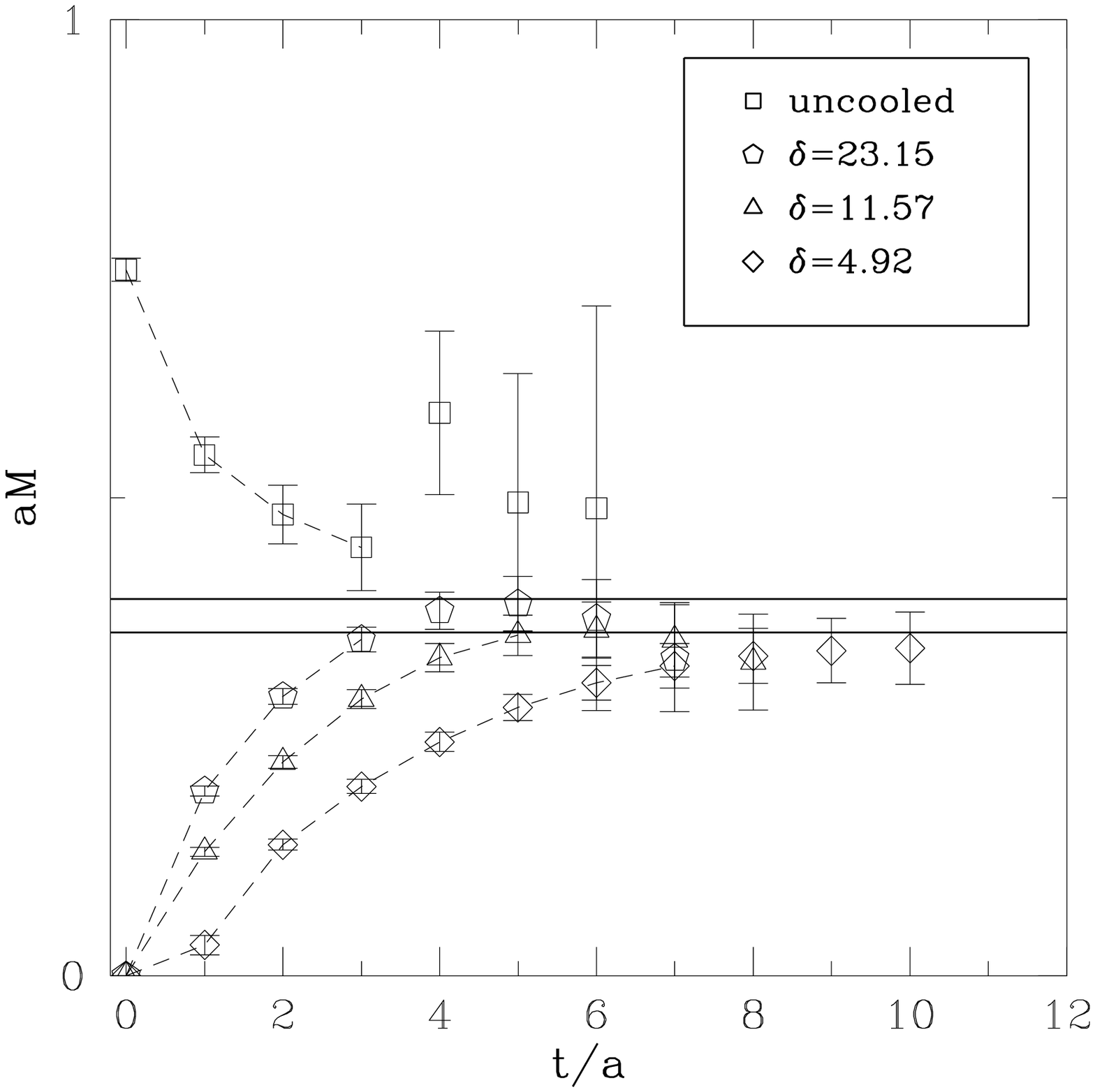} 
\includegraphics{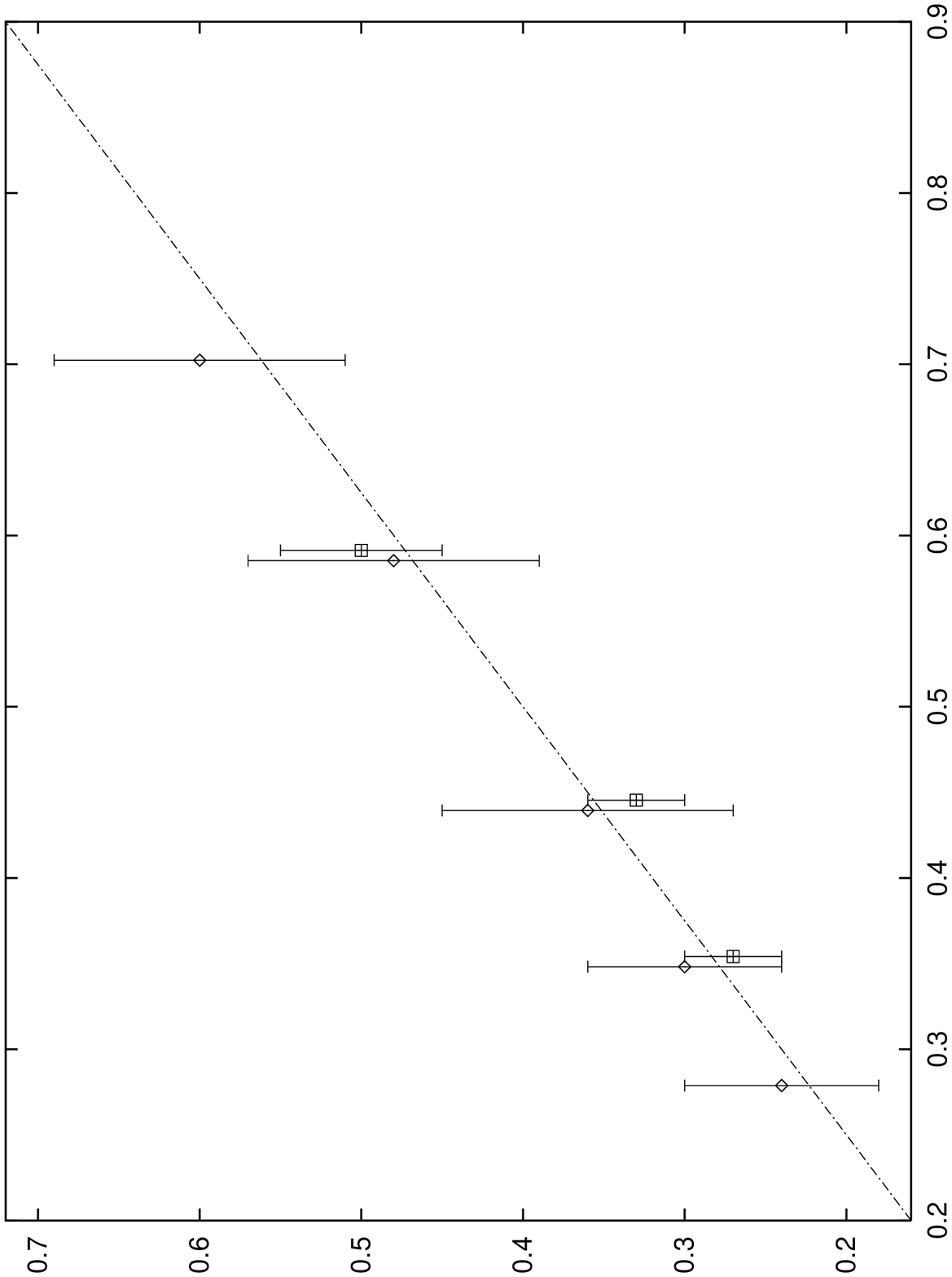}
\caption
{Left and middle: $M(t)$ for the $a=0.12$ ($0.06$) fm lattice [5]. 
Horizontal bands represent 
standard results for $\sigma$. Right:
The smoothing scale of RIC $r_c(\delta)$ vs $\delta^{-1/3}$ 
for the $a=0.12$ fm
(diamonds) and $a=0.06$ fm (squares) lattices. Also
shown is the
fit (\ref{e.rde}).}
\label{f.rde}
\vspace{-0.4cm}
\end{figure}

We present in Fig. \ref{f.rde} $M(t)$ for $SU(2)$.
Defining  $r_c(\delta)$
as the distance $t$ at which $M(t)$ on $\delta$-cooled configurations
starts to agree with the uncooled value (obtained by fuzzing and fitted 
to a smooth function of $t$) we arrive at  
the results in Fig. \ref{f.rde} (right) showing the scaling behaviour
 expected from the continuum limit of the parameter  $\delta$, 
compatible with
\be
r_c(\delta) \simeq 0.8 \ \delta^{-1/3} .  
\label{e.rde}
\ee

The behaviour of instantons is similar under RIC and IC. RIC
has a  dislocation threshold depending on $a^3\delta$,
$\rho_0(a^3\delta) = {\hat \rho}_0(a^3\delta) a$ approaching the 
IC dislocation threshold from below: ${\hat \rho}_0(a^3\delta) \leq
{\hat \rho}_0(0) \equiv {\hat \rho}_0 =2.3$. 
Fixing $\delta$ and rescaling $\beta$ to approach the
continuum limit the RIC threshold shrinks similarly to the IC one.
 Instantons
above the dislocation threshold are preserved unchanged both by RIC
(independently on $r_c$) and by IC (independently on cooling sweeps),
since both algorithms use the same action with scale invariant
instanton solutions.
I-A pairs, however, do not annihilate under RIC the way they 
do for the other cooling methods (including IC): 
since the distortion of the 
partners in a pair depends on the overlap,
RIC preserves pairs below some overlap-threshold. More precisely, 
 there is a well defined 
relation between $\Delta_{\mu}$ (measured at the center of the partners
or at mid-distance between them)
and $S_{\rm int}^{\rm IA}$ eq. (\ref{e.over}).
 Thus RIC   
stabilizes I-A pairs with $S_{\rm int}^{\rm IA}$ below  a 
threshold  depending on  $r_c(\delta)$.

\section{Topological Properties of Yang-Mills Theories at $T=0$}

\subsection{Global Properties [1],[3],[5],[6]}

The topological charge given by IC is already after a few 
cooling sweeps 
 an integer to about $1\%$ and stable thereafter. 
Similarly, $Q$ given by RIC is also an integer 
and independent on $\delta$ if the cooling radius is larger than
the dislocation threshold, $r_c > \rho_0 \sim 2a$. The topological
charge distribution is therefore  well defined. It also agrees
excellently, e.\,g.,  with the distribution obtained by using overlap fermions
\cite{nara}. The topological susceptibility obtained in this way
is well defined, shows good scaling properties and agrees with the
Witten-Veneziano relation. See Table \ref{t.sus}.

\begin{table}[h]
\begin{center}
\begin{tabular}{|c|c|c|c|c|c|c|c|} 
\hline
 & $SU(2)$& $SU(2)$& $SU(3)$& $SU(3)$& 
$QCD$& $QCD$& $QCD$ \\
\hline
Latt. & $12^336$& $24^4$& $12^4$& $16^4$& 
$24^312$ & $24^312$&$24^312$ \\
\hline
bound. cond. &  pbc& tbc& tbc& tbc& 
apbc& apbc& apbc \\
\hline
$a$ (fm)& 0.12 & 0.06 & 0.134 & 0.1 &
0.115& 0.103 & 0.0855 \\ 
\hline
$\chi^{1/4}$ (MeV)& 195(4) & 200(8) & 184(6) & 182(7) & 134(10) &
102(5) & 0 \\
\hline

\end{tabular}
\caption{$\chi^{1/4}$ for pure Yang-Mills  at $T=0$ [1],[3],[5] (average)
 and for 
QCD at $T\,\simeq\,0.85,\,0.98$ and $1.13T_c$ [3]. The results are
given for $n_c>20$ or $r_c>2a$.\label{t.sus}
}
\end{center}
\vspace{-0.6cm}
\end{table}

\subsection{Local Properties [1],[3],[5],[6],[7]}

\subsubsection{Description of the I-A Ensemble and Overview of the Results}

In extracting the instanton information
out of the configurations we approximate the action and charge
density by a superposition of self-dual, non-interacting I's and 
A's parameterized through the 1-instanton BPST ansatz (\ref{e.dens}) 
(partially corrected for periodicity by adding images).
We measure the departure of the real action and charge density from 
the above non-interacting ansatz through the quantities
\be
\varepsilon_s = \sqrt{
{{\int d^4x |s(x) - s_{\rm fit}(x)|^2} \over {\int d^4x |s(x)|^2}}},\ \ 
\varepsilon_q = \sqrt{
{{\int d^4x |q(x) - q_{\rm fit}(x)|^2} \over {\int d^4x |q(x)|^2}}}.
\label{e.dfit}
\ee
\no In the actual proceeding every time an instanton candidate
is located, it is counted only if, by adding it to $s_{\rm fit}$
and $q_{\rm fit}$, $\varepsilon_{s,q}$ simultaneously decrease.
 Of course this description misses more complicated
topological structures (e.g., fractional charge) -- it is a question of
self-consistency  to apply it within a picture of the
instanton ensemble consisting of typical,  self-dual I's and A's in
weakly interacting superpositions. 
Attempts to use a more general description have been made in the literature 
\cite{michael}, their theoretical understanding appears to us difficult, however.

This analysis is performed at a given degree of smoothness and we ask about
the stability of the results with the latter. Our attitude is that significant
dependency on the smoothing degree can reflect, may be unwanted, but real 
properties of 
the instanton ensemble. To judge about this scaling tests are essential. 
Normal cooling and even IC reveal the smoothing 
dependence as  dependence on the 
number of cooling steps. 
The latter, however, has only an indirect physical meaning \cite{tep1}. It
  needs to be
calibrated in some way with $\beta$ (see \cite{mnpnp},\cite{smtep} and for a
recent systematic attempt \cite{ringw}), the criterion for the
calibration remaining, however, just the property which one wants to prove, 
namely scaling. RIC, on the other hand, has an automatic, 
natural calibration via the
physical (dimensionfull) cooling scale $r_c$. The cooling is 
independent on $\beta$ if
 $r_c$ is larger than the shrinking dislocation threshold $\rho_0$,
and therefore provides a genuine
scaling check \cite{ric}. We then consider 
any $r_c$ dependence which subsists the scaling test as potentially 
relevant and do not attempt 
a further treatment of the data (e.g., 
extrapolation to 0 smoothing degree \cite{boul},\cite{neg}) since we
are not clear about the physical basis of the dependence and of the treatment.

An overview of the   $T=0$ RIC and IC $SU(2)$ results
is given in Table \ref{t.su2}, for $SU(3)$ and $QCD$ we refer to \cite{buck}. 

\begin{table}[ht]
\begin{center}
\begin{tabular}{|c|c|c|c|c|c|c|c|c|}
\hline
a &$\delta$
& $r_c(\delta)$ 
& $\langle N/V \rangle$
& $\langle |Q|\rangle$
& $\langle d_{II}\rangle$
& $\langle d_{IA}\rangle$
& $\varepsilon_q$
 ($\varepsilon_s$)
& $\langle \rho\rangle$
\\
\hline
\hline
$0.12$&23.15  & 0.30(5) &14.86(7)&2.8(1)&0.39&0.28&0.75(0.55)&0.38\\ \hline
"           &11.57  & 0.35(5) &8.72(2)&2.93(9)&0.46&0.36&0.60(0.45)&0.40\\ \hline
"           &4.92   & 0.48(5) &4.21(2)&2.91(6)&0.54&0.49&0.45(0.35)&0.42    \\ \hline
"           &2.89   & 0.60(5) &2.70(3)&2.85(8)&0.59&0.58&0.35(0.30)&0.44    \\ \hline
"           &IC(20) &        &2.06(2)&2.91(8)&0.61&0.61&0.30(0.30)&0.42    \\ \hline
\hline
$0.06$  &23.15  & 0.27(3) &18.7(1)&1.7(1)&0.40&0.30&0.60(0.45)&0.35\\ \hline
"           &11.57  & 0.33(3) &9.6(1) &1.7(1)&0.48&0.39&0.45(0.40)&0.38\\ \hline
"           &4.92   & 0.50(3) &4.35(6)&1.8(1)&0.57&0.53&0.25(0.25)&0.40\\ \hline
"           &IC(50) &        &2.96(5)&1.7(1)&0.59&0.57&0.25(0.25)&0.37\\ \hline
\end{tabular}
\caption[]{\label{t.su2}
{$SU(2)$  results for IC at 20 and 50 cooling sweeps
and for RIC at various $\delta$ [5] -- see also Table \ref{t.sus}. 
The lattice spacing ($a$), the 
cooling radius ($r_c(\delta)$), 
distances and sizes are given in fm.
The density $\langle N/V \rangle$ is in fm$^{-4}$ 
and $\delta$ in fm$^{-3}$. The errors are statistical only,
when they are not explicitly quoted they amount to about 1 
in the last indicated digit. 
}}
\end{center}
\vspace{-0.4cm}
\end{table}

\subsubsection{Instanton Size Distributions}

\begin{figure}[htb]
\vspace{7cm}
\includegraphics{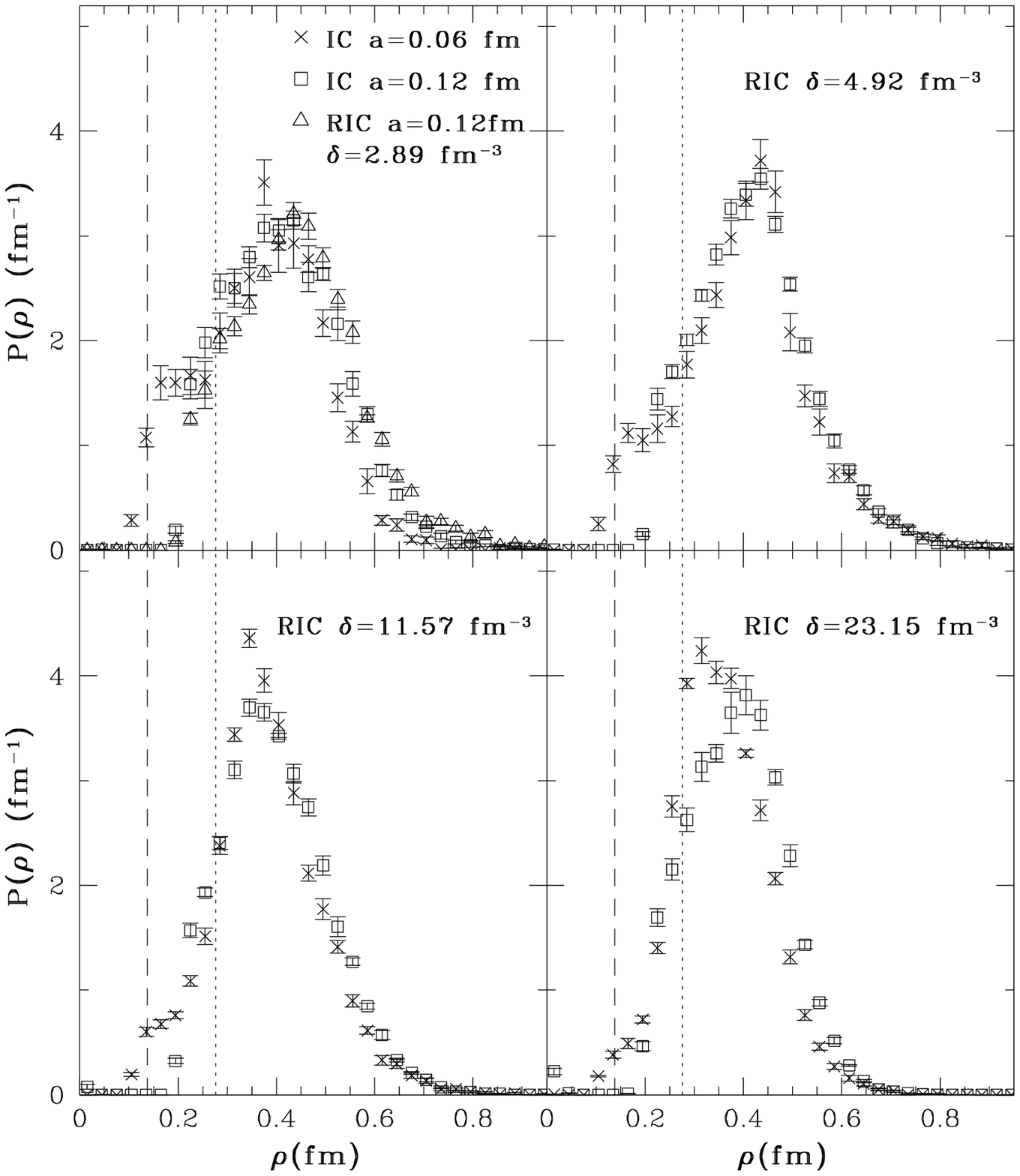}
\includegraphics{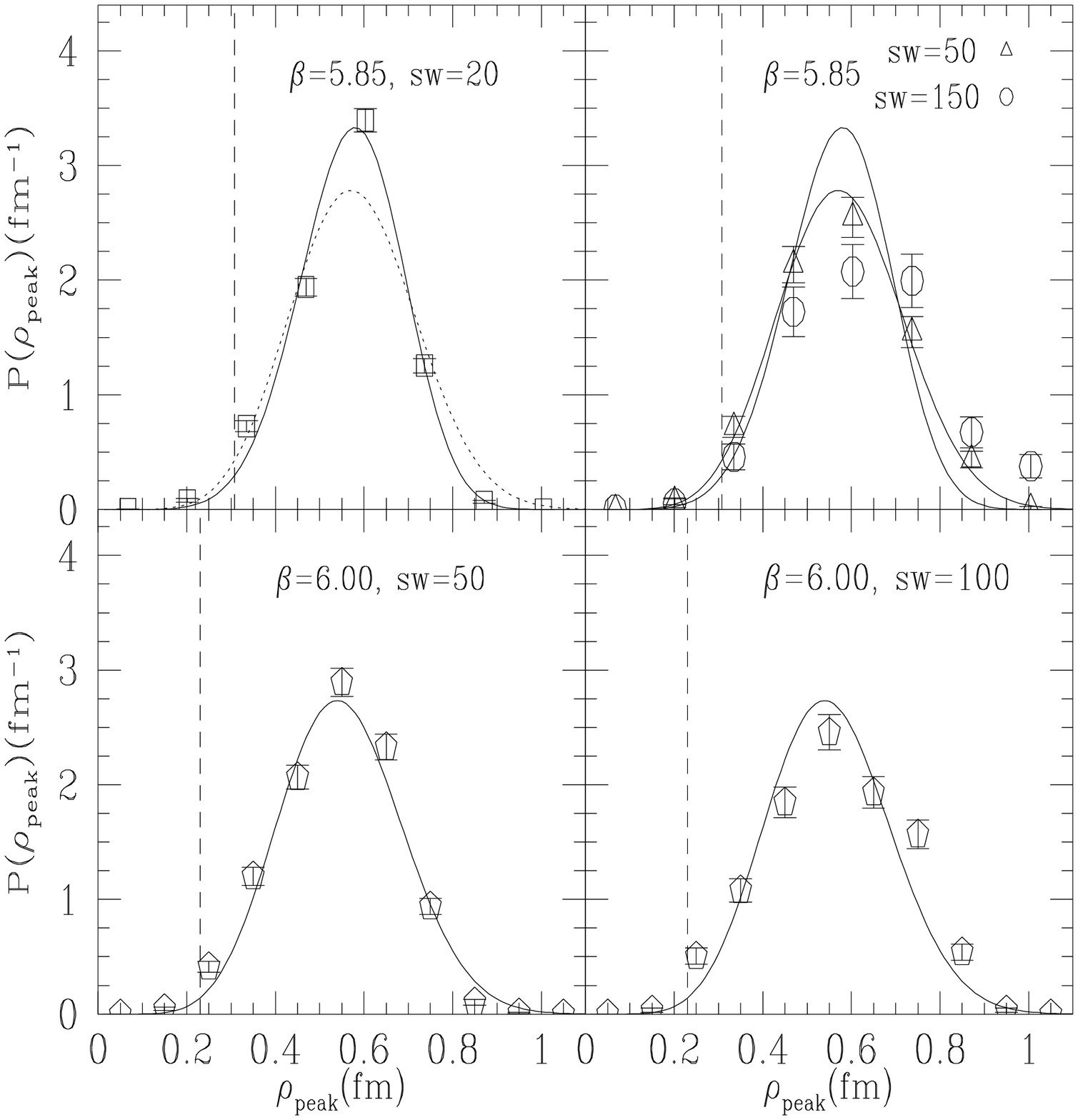}
\caption{Left: Normalized size 
distributions for $SU(2)$ from RIC and IC
on the $12^336$, pbc, $a=0.12$ fm lattice (squares, triangles)
and on the $24^4$, tbc, $a=0.06$ fm lattice (crosses) [5]. The RIC data correspond to
$r_c\, \sim \, 0.56,\, 0.47,\, 0.35$ and $0.28$ fm ($\delta\, =\, 2.89,
\, 4.92,\, 11.57$ and $23.15$ fm$^{-3}$, respectively).
Vertical dotted (dashed) lines indicate the  IC dislocation
threshold $\rho_0=2.3a$ for $a=0.12$ ($0.06$) fm. The corresponding
RIC dislocation thresholds are lower. Right: Normalized size 
distributions for $SU(3)$ from IC
on the $12^4$, tbc, $a=0.134$ fm lattice (squares, circles, triangles)
and on the $16^4$, tbc, $a=0.1$ fm lattice (octagons) [3]. The curves are Gaussian fits, the dashed lines $\rho_0=2.3a$.}
\label{f.size2}
\vspace{-0.4cm}
\end{figure}

\begin{figure}[htb]
\vspace{8.cm}
\includegraphics{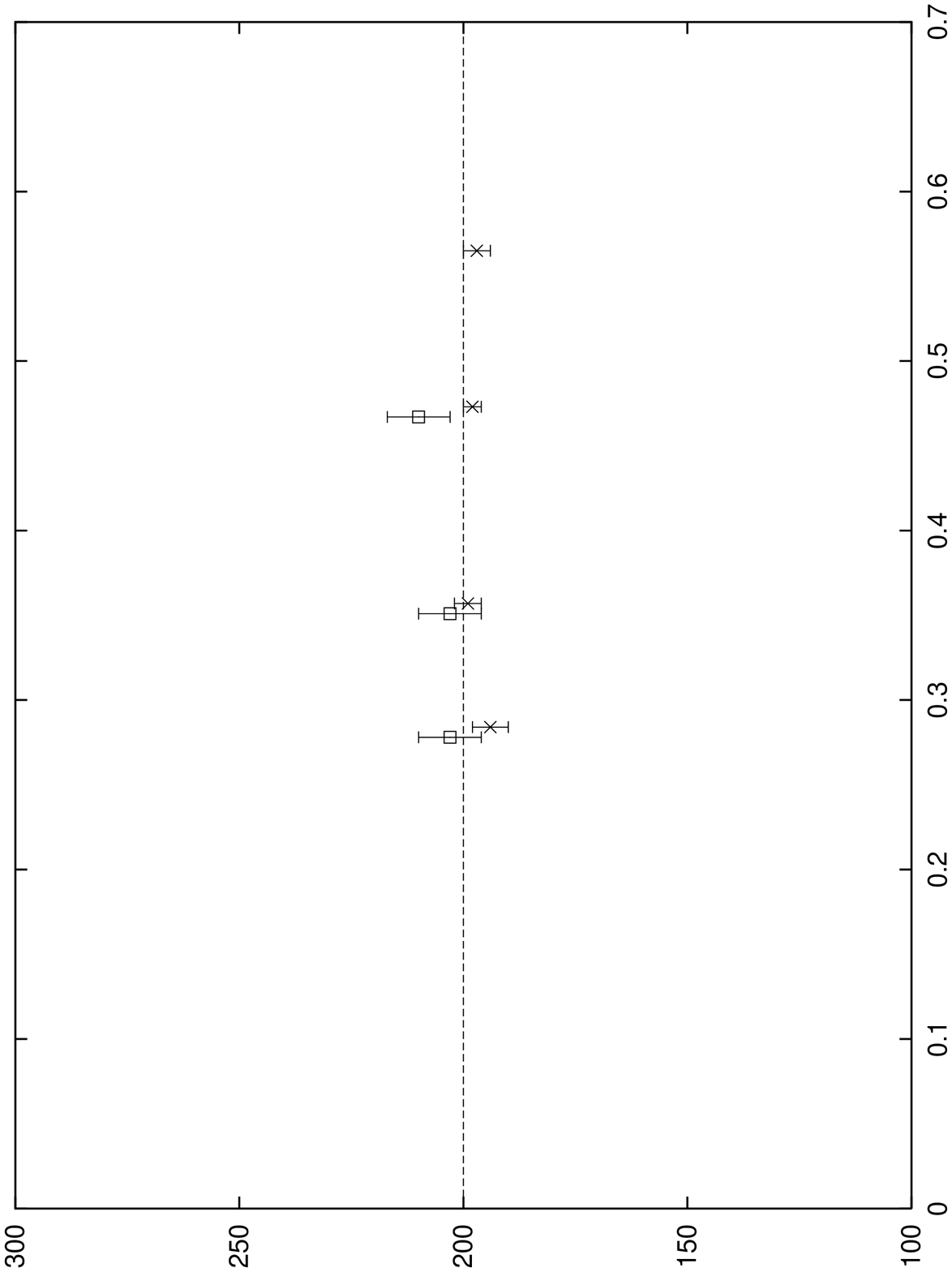}
\includegraphics{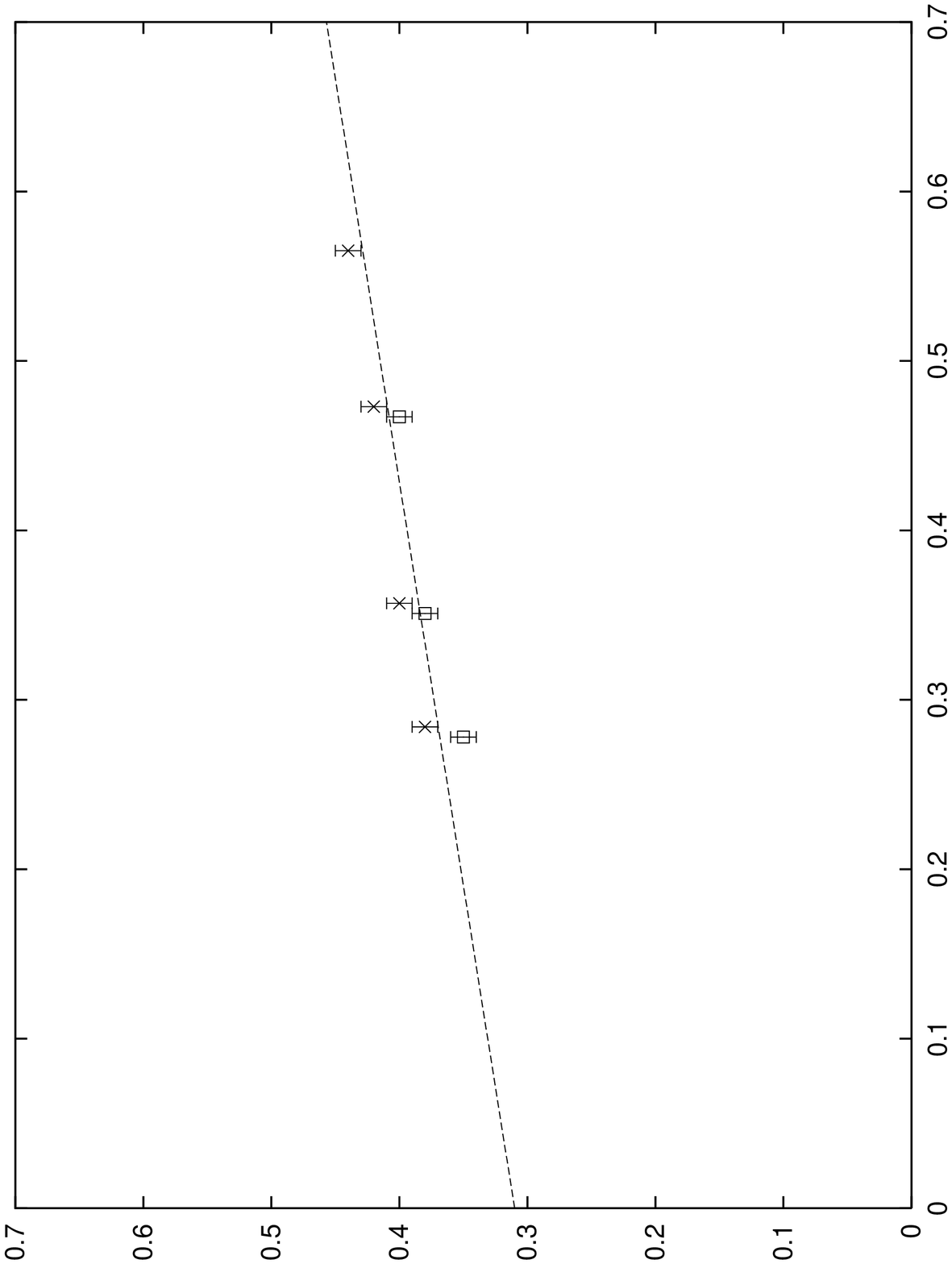}
\includegraphics{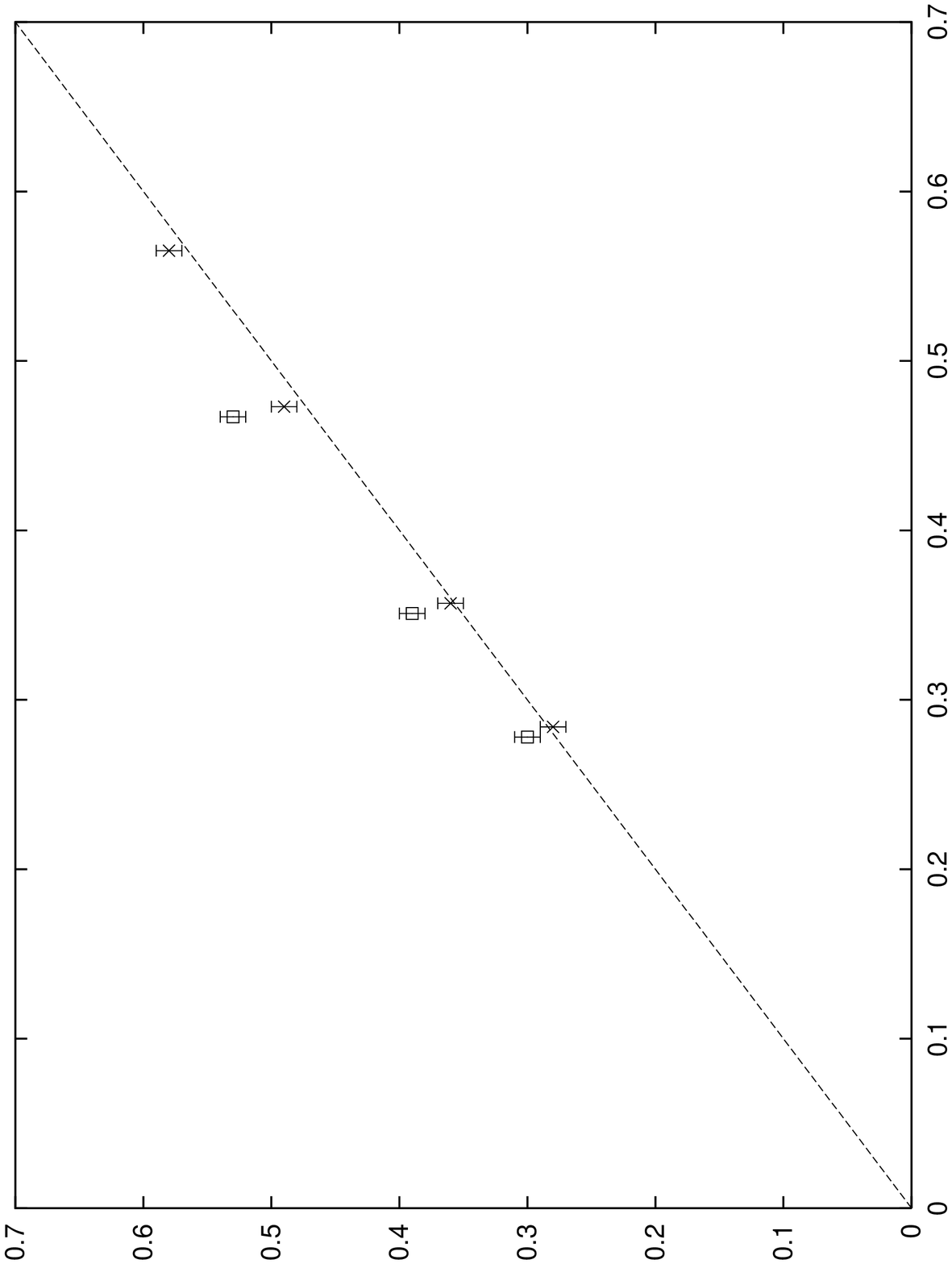}
\includegraphics{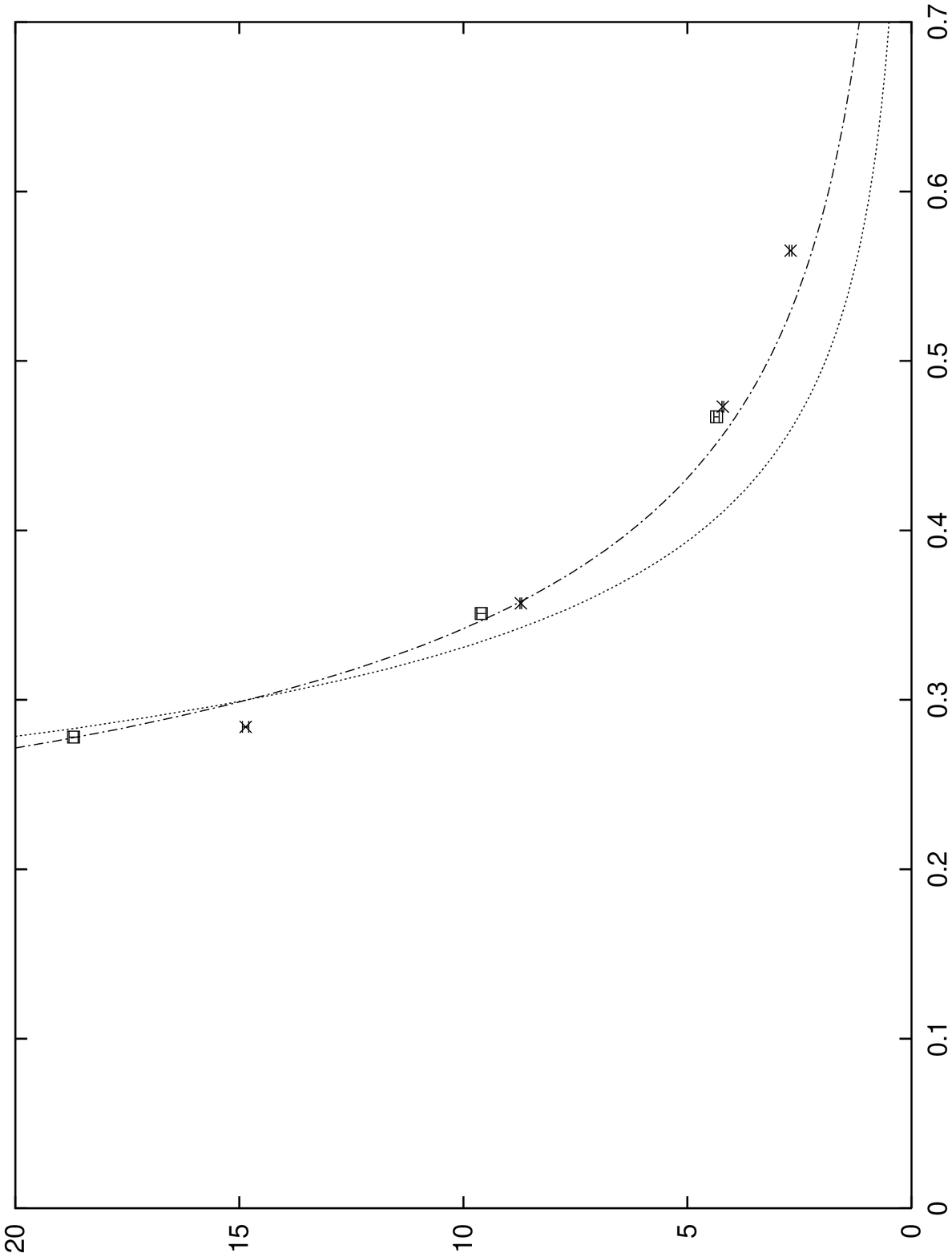}
\caption{Smoothing scale  dependence of: $\chi^{1/4}$ (upper left), 
average I(A) size (upper right), average I-A distance, (lower left) and 
average density 
(lower right) for $SU(2)$ 
at $a= 0.12$ fm (crosses) and $a=0.06$ fm (squares) [5]. $r_c$ is taken 
from the fit (\ref{e.rde}), the horizontal
errors (not shown) are about 0.04 fm -- see also Table \ref{t.su2}. The  lines 
illustrate linear (dashed), $r_c^{-3}$ (dashed-dotted) or
$r_c^{-4}$ (dotted) behaviour.}
\label{f.extra}
\vspace{-0.4cm}
\end{figure}

As can be seen from Fig. \ref{f.size2} the size distributions show remarkable 
stability under changing  the RIC smoothing scale and under rescaling of the
lattice spacing, up to a limited variation of the position of the maximum 
(notice that these are
normalized distributions, the number of instantons varies by more
than a factor 5 between small and large $r_c$).
Also the IC data show the same stability, both for $SU(2)$ and for $SU(3)$ --
see Fig. \ref{f.size2}. The distributions are little 
(at larger $a$) or not affected (at smaller $a$) by the dislocation threshold.
The average sizes $\rho_{av}$ 
are $0.4 \pm 0.04$ fm for $SU(2)$ and $0.55 \pm 0.06$ fm for 
$SU(3)$,
with full widths of about $0.2\,(0.25)$ fm, respectively. Since 
the variation of $\rho_{av}$, which may
have different origins \cite{PMsiz}, is rather limited  we
are tempted to consider this quantity as well defined (inside errors accounting
 for the various uncertainties) -- see also Fig. \ref{f.extra}.
The small size branch of the distribution grows more slowly than the power law 
 of the dense gas  approximation (a fit gives $p = 1.2 \pm 0.5$  for $SU(2)$ and  $p = 2.5 \pm 1.0$
 for $SU(3)$,  instead of 
 $p_{\rm dilute} = 2.33$ and $p_{\rm dilute} = 6$, respectively) 
which indicates that in the region of small but relevant sizes the
  dilute gas picture is not adequate. The large size behaviour of the
 distributions  cannot be determined from these simulations 
since instantons 
 of size significantly above 1/3  of the lattice can show finite 
lattice size effects and
 already $\rho_{av}$ is near this point ($0.48$fm for $SU(2)$  
and $0.53$ fm for $SU(3)$; but the existence and position of the
maximum are safe from this point of view).

\begin{figure}[htb]
\vspace{4.5cm}
\includegraphics{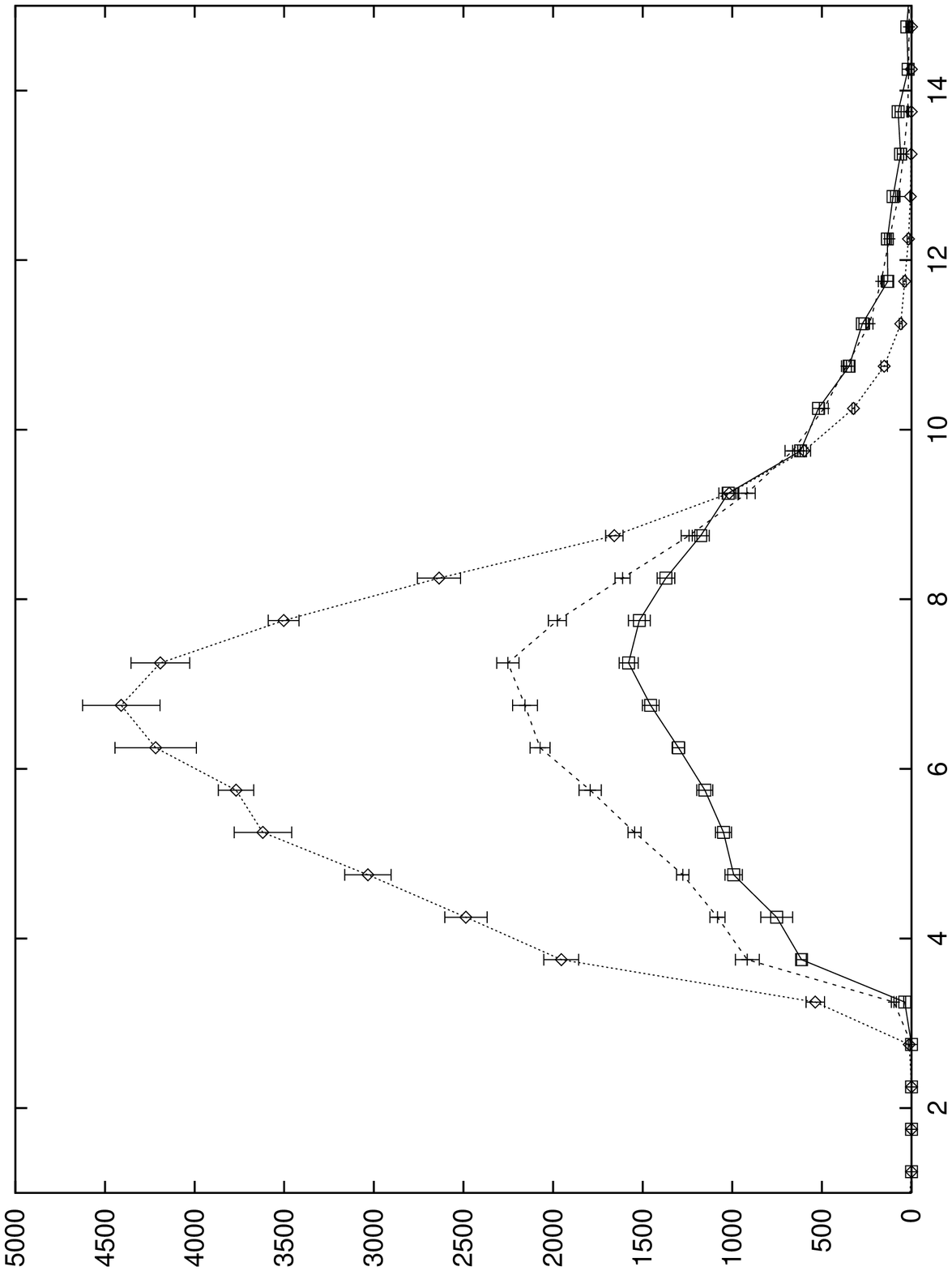}
\includegraphics{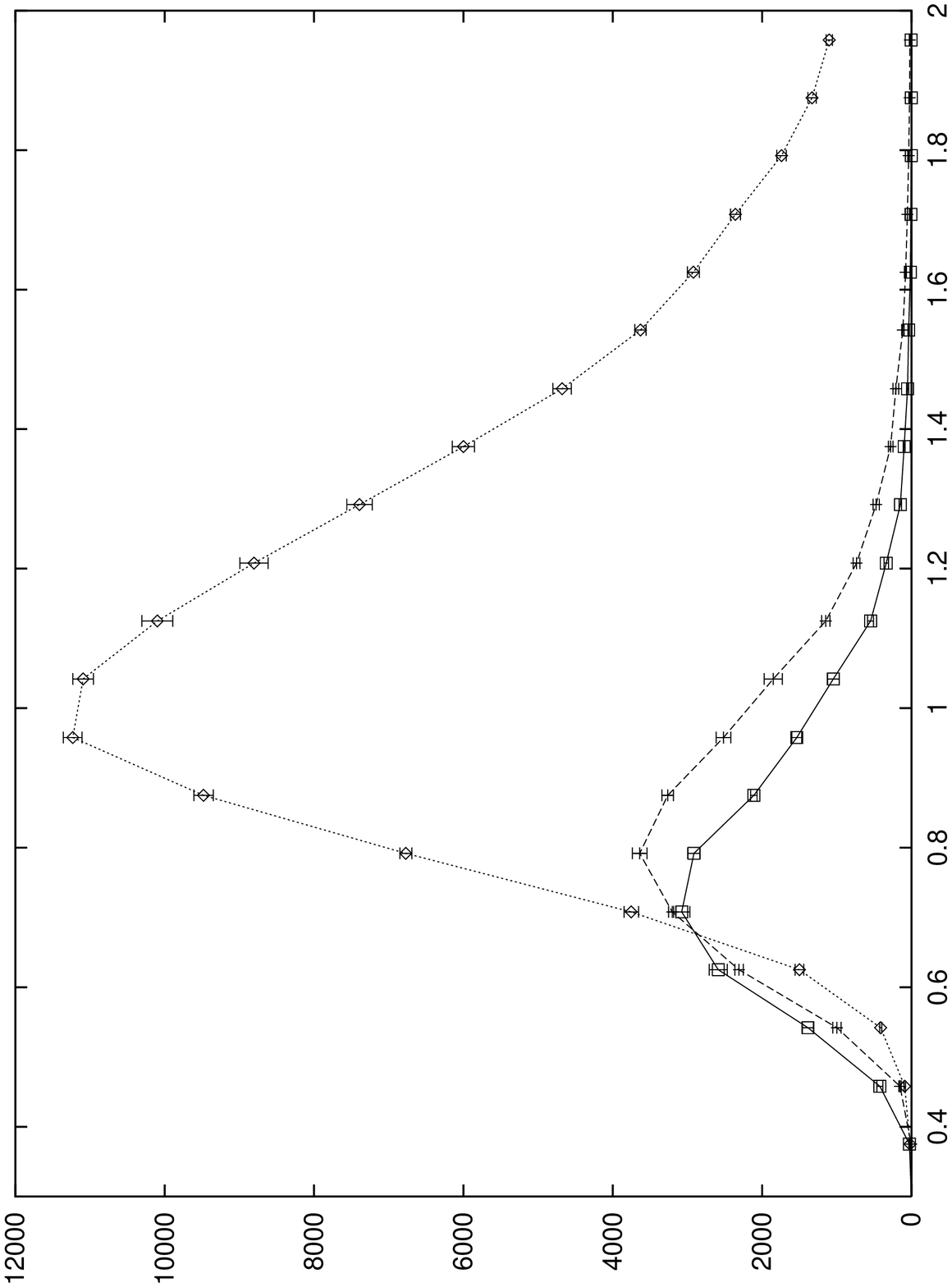}
\caption{Non-normalized size
and overlap distributions from RIC
 at $a=0.12$ fm [5] for 
smoothing scales $r_c\, \sim 0.56$ (squres), $\sim 0.47$ (pluses) 
and $\sim 0.28$ (diamonds) fm. The scale for sizes is  $0.06$ fm.}
\label{f.ov}
\vspace{-0.4cm}
\end{figure}

\subsubsection{Instanton Density and Pair Properties}

The number $N$ of I's and A's observed
at small cooling radius is essentially due to pairs ($|Q|$
is constant and relatively small) -- see Table \ref{t.su2}.
We notice that while $N$ increases by roughly a factor  5 when
$r_c$ is decreased from about $0.6$ fm  to about $0.25$ fm, $\rho_{av}$
only varies by about $10\%$. Consequently, the packing fraction $f$ (\ref{e.pac})
also increases approaching 1, which indicates that at small $r_c$ we
observe a rather dense ensemble, strongly deviating from  Poisson distribution
($c_{\rm P}$ eq. (\ref{e.corr}) is about 0.5) and with a tendency 
to charge neutralization in small regions (see \cite{mnpnp},\cite{smtep}).
The overlap (\ref{e.over}) distribution shifts toward larger values 
\cite{latt99} while the average  average distance of an I to the nearest 
A ($d_{IA}$) decreases.
 See Fig. \ref{f.ov}.
The situation is succinctly 
described in Fig. \ref{f.extra}. Here we plot
 $SU(2)$ RIC data for averages
from two lattices ($a=0.12$ fm and $a=0.06$ fm) at given $r_c = 0.8
\delta^{-1/3}$, see eq. (\ref{e.rde}), i.e., 
the dimensionless $\delta a^{-3}$ is simply rescaled with $a$ at fixed $\delta$.
As already noticed, 
we see that the susceptibility 
$\chi$ and the average size $\rho_{av}$ scale well with the
cut-off and show no ($\chi$) or little dependence ($\rho_{av}$)
on the smoothing scale. Also the average I-A distance 
 and the density of I(A)'s have comparable  
 values on the two lattices and seem therefore to scale. However, their
dependence  on  $r_c$ is 
disturbing: the instanton density increases as an inverse power of $r_c$
and the average I-A distance
is practically given by $r_c$ itself. Simultaneously with increasing density also 
the quality of the fit decreases ($\varepsilon_q ,\, \varepsilon_s$ increase,
see Table \ref{t.su2}). This suggests that the structure revealed at small 
smoothing scale (large frequencies) is strongly overlapping and distorted.
Generally it appears  
that a ``typical" density or I-A distance could only be defined by first
fixing the smoothing scale. By contrast, the average size seems to have a well
defined meaning for itself. Finally, restricting smoothing to scales below
0.25 fm retains too much short range fluctuations. 
 We stress that the cut-off rescaling of these results is done 
without any ad-hoc fit of cooling parameters.\footnote{The small, systematic 
difference of a few
percents between the ``$a = 0.12$ fm" and the ``$a=0.06$ fm" data 
for all observables (including $\sigma$ and $\chi$) \cite{ric} may be due to 
a minimal uncertainty in assessing the value of $a$ \cite{mnpnp} as taken 
from the literature.} 

\section{Topology at $T>0$ and the Chiral Connection [2],[3],[4],[7]}

\begin{figure}[htb]
\vspace{4.8cm}
\includegraphics{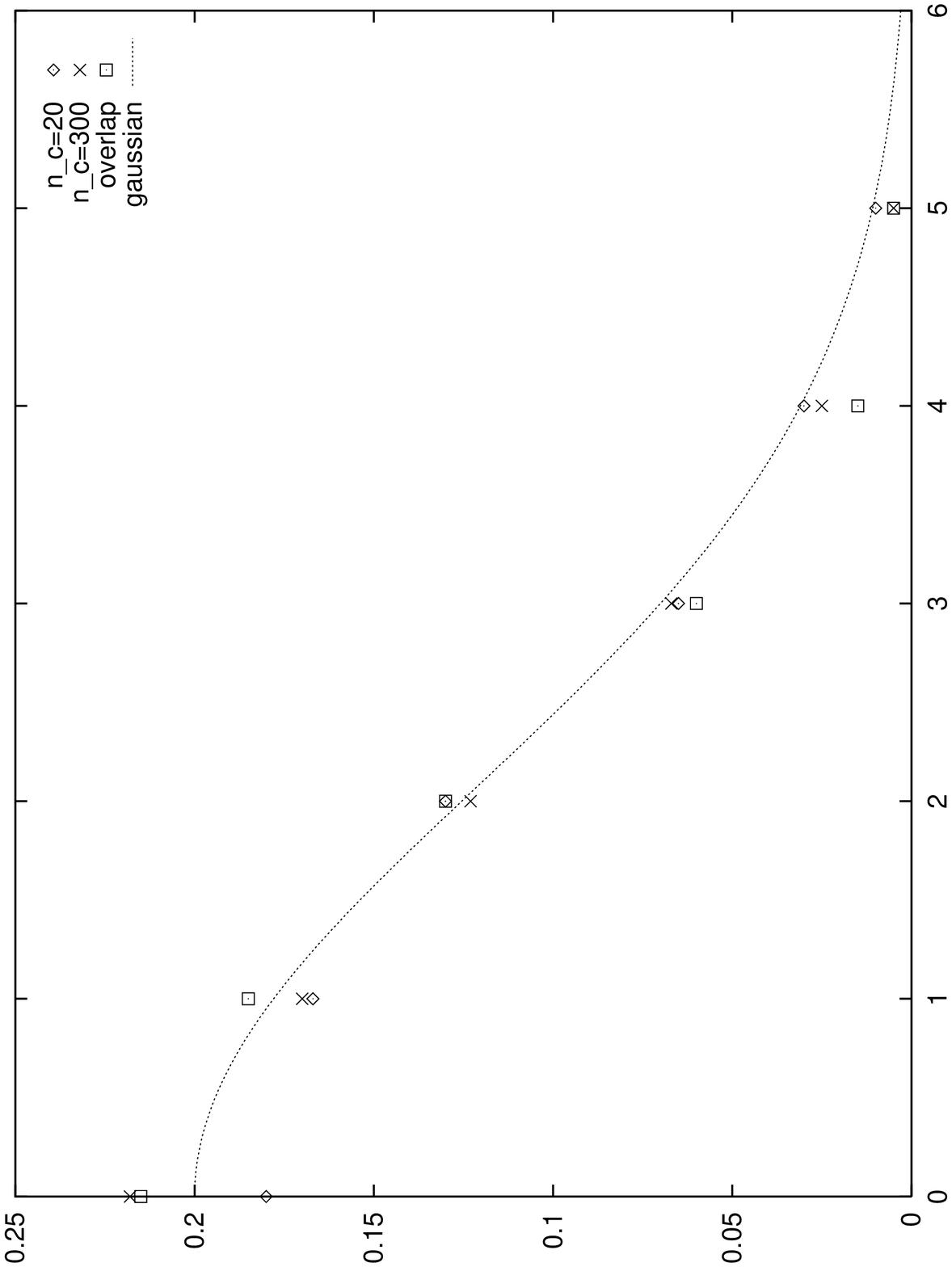}
\includegraphics{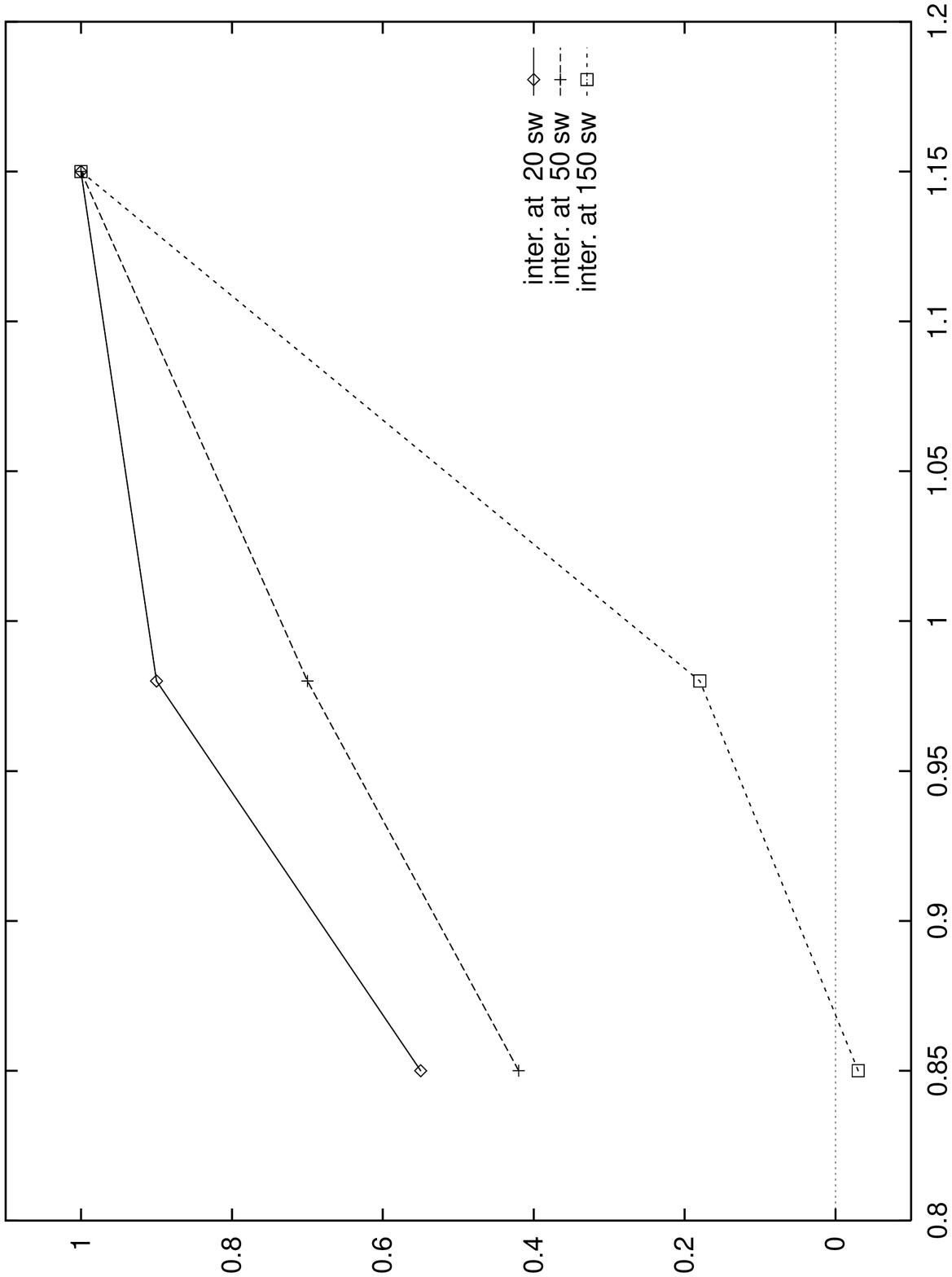}
\caption{Left: Topological charge distribution for pure $SU(2)$ by IC at
$n_c=20$ (diamonds) and $n_c=300$ (crosses) and via overlap formalism (squares). The curve is an illustrative Gaussian with arbitrary normalization and slope corresponding to
a susceptibility of $(200$ MeV$)^4$. 
Right: Temperature dependence of the full $QCD$ data of Fig. \ref{f.milc}: 
$c_{\rm int}$ for $T\,\sim\, 0.85,
\,0.98$ and $1.13T_c$
after 20 (upper data), 50 (middle) and 150 (lower data) 
IC cooling sweeps.}
\label{f.chir}
\vspace{-0.4cm}
\end{figure}

A first impression on the connection between topology and chiral properties 
 (see \cite{ada} for a recent review) is obtained at $T=0$, e.\,g.,  by using 
the definition of the topological charge via the Ward identities for 
Wilson fermions \cite{entop} or via the overlap formalism \cite{nara} and comparing with the improved topological charge measured
by IC or RIC. In Fig. \ref{f.chir}, left plot, we present the 
topological charge distribution measured on a $12^4$, $SU(2)$ lattice
($a=0.12$ fm, pbc \cite{mnpnp}) by IC and by overlap formalism \cite{nara}. 
See also  \cite{schi}.

  Increasing the temperature above $T_c$ produces radical changes in the topological
properties: the susceptibility drops and the instanton ensemble seems to
change its character \cite{pisat},\cite{buck},\cite{latt97}. 
Above $T_c$ practically all configurations are in the
$Q=0$ sector although at small $r_c$ RIC reveals a quite large $N$ (a similar
thing is observed with IC at small number of cooling sweeps). 
The ensemble 
consists of nearby opposite charges, but it is difficult to say if these
form ``molecules" or just a strongly fluctuating background. 
The description in terms of I's and A' becomes difficult (the fit  
deteriorates). On Fig. \ref{f.chir}, right plot, we show the I-A correlation
parameter 
$c_{\rm int}$ (\ref{e.cint}) for the full QCD, MILC configurations below and above $T_c$ \cite{milc}.  

\begin{figure}[htb]
\vspace{4.8cm}

\includegraphics{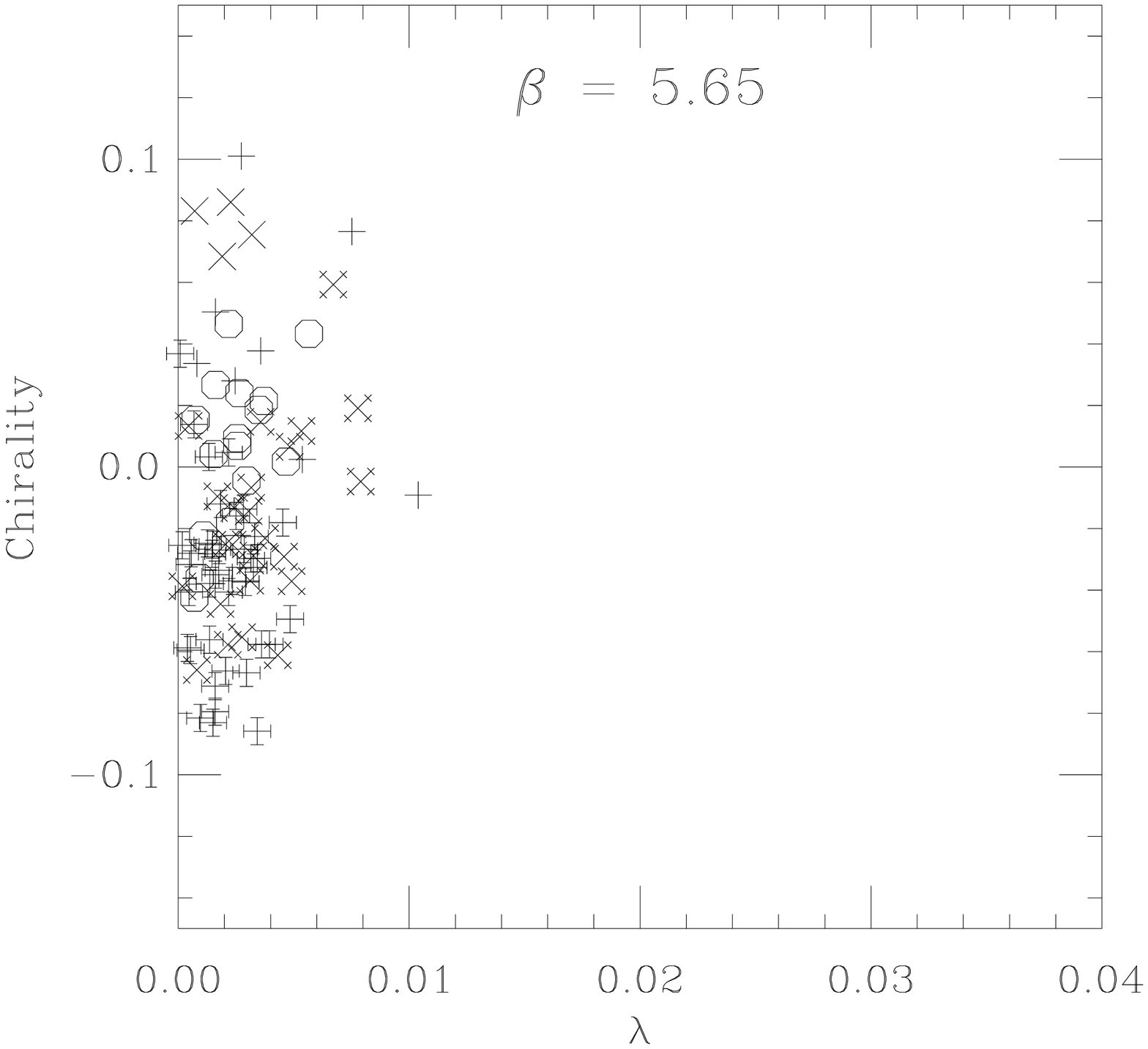}
\includegraphics{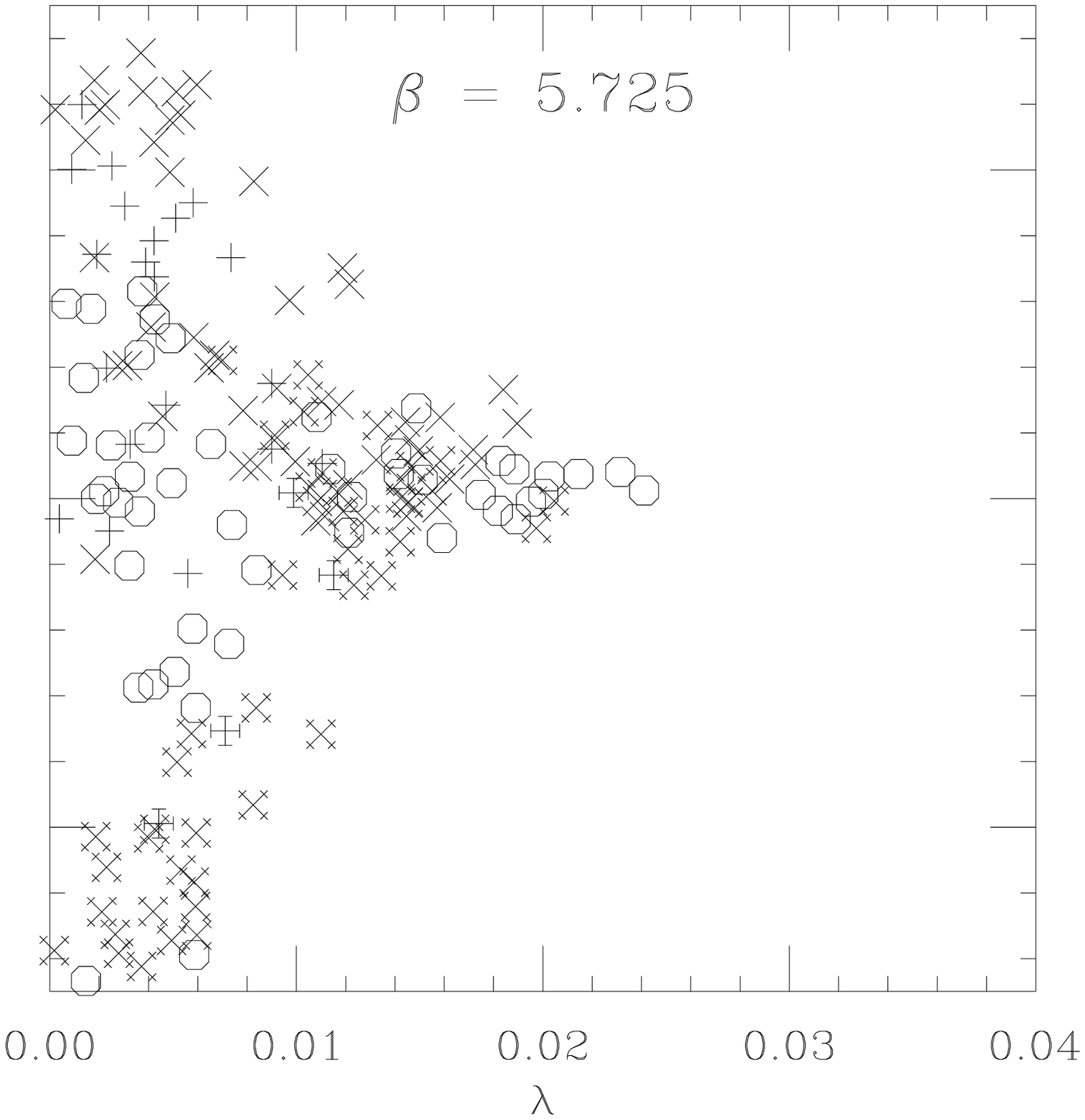}
\includegraphics{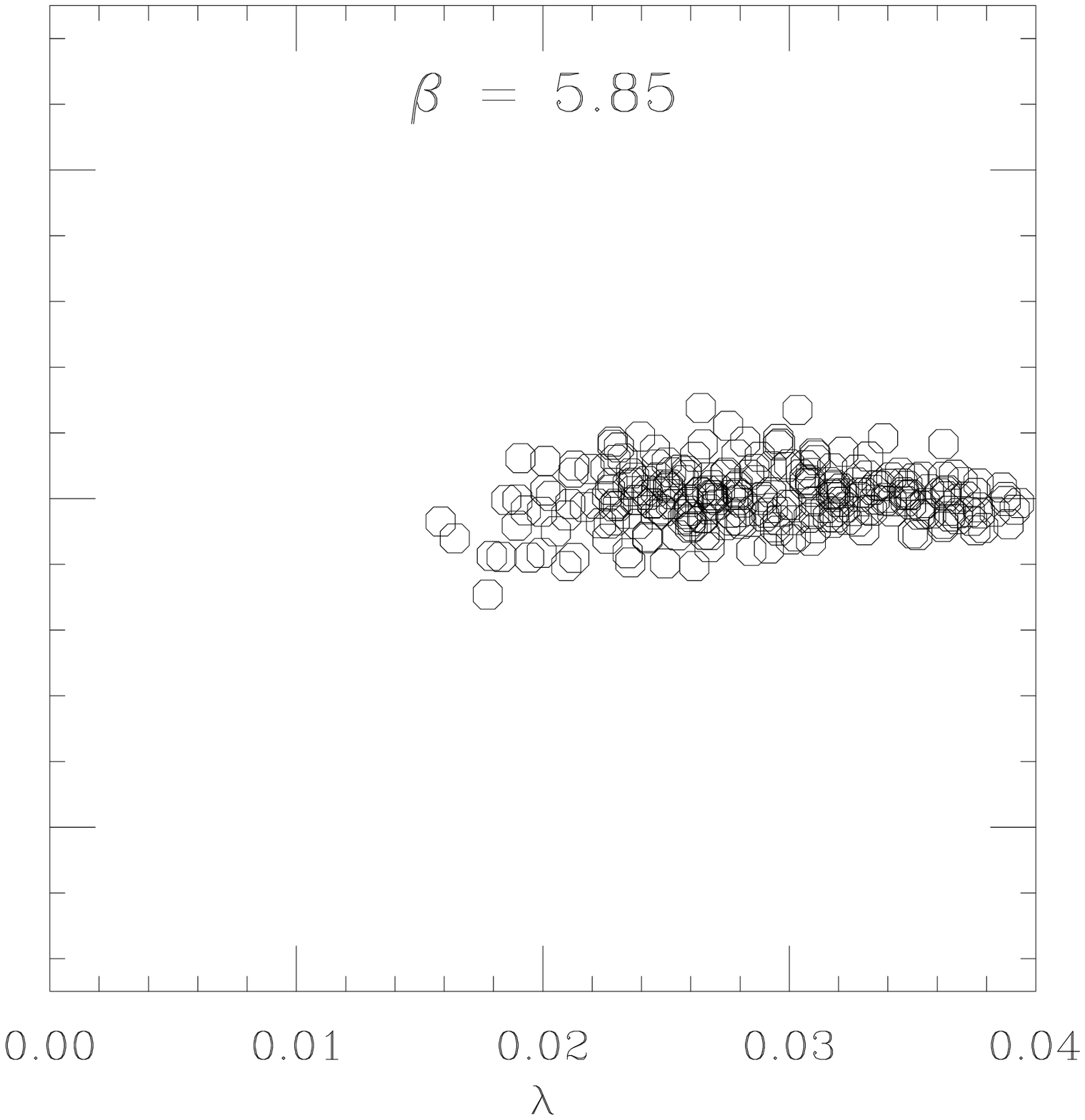}
\caption{$QCD$ with dynamical quarks: Chirality/eigenvalue plots for the lowest 4  
eigenvalues
using symbols corresponding to the topological charge for each configuration:
octagon ($Q=0$), cross and fancy cross ($Q=\pm 1$) and
plus and fancy plus ($Q=\pm 2$). The three plots correspond to $T\,\sim\, 0.85,
\,0.98$ and $1.13T_c$ [7]}
\label{f.milc}
\vspace{-0.4cm}
\end{figure}

It is interesting to observe the correlation between topological and chiral 
properties in full QCD as function of the temperature. On the MILC  configurations \cite{milc} which use two
flavours of staggered quarks we have measured the improved 
topological charge and the 
chirality of the lowest eigenvalues of the Dirac operator. Although the
renormalization effects on the chirality are large there appears to be a
strong correlation between the total chirality carried by the small eigenvalues 
(unsmoothed)
and the total charge of the configuration obtained by IC. Both are also similarly
affected  by
the temperature: above $T_c$ $Q=0$ and all Dirac modes become heavy and have
vanishing chirality. See Fig. \ref{f.milc} where also the  I-A correlation
parameter 
$c_{\rm int}$ (\ref{e.cint}) is drawn.

One even  observes local correlation between instantons and Dirac modes
\cite{latt97},\cite{latt98}, the bulks of the corresponding densities seem
to overlap quite well.

\section{Conclusions}

The question: which are the vacuum excitations relevant for the dynamics of
Yang Mills theories? can be studied in the frame of
Euclidean lattice theory by 
observing the structure of the configurations generated in
Monte Carlo simulations.
Besides instantons various other proposals have been made, such as
 super-instantons \cite{erpa}, abelian monopoles or vortices 
(for recent reviews see, e.\,g., \cite{ADG},\cite{vort},\cite{all}).
 We have here been concerned with 
topological properties, that is with the problem of observing 
and characterizing the (anti-)instanton ensemble in the MC configurations.
Thereby the I's and A's are identified as self-dual structures and fitted
to the BPST ansatz. We obtain in this way various results (density, charge and size distributions etc) 
and we also check the scaling behaviour of these properties 
under changing of the lattice spacing by factors of the order of 2.
In agreement with other studies (see \cite{neg},\cite{tepr} for recent reviews), our results indicate that 
global observables like the 
topological susceptibility and charge distributions appear well
defined and show good scaling properties. Also some 
local properties, like size distributions of instantons appear
reasonably well defined and scaling. It seems, however, that the
topological charge density can fluctuate strongly at small distances and
a description of these fluctuations in terms of instantons and anti-instantons
is difficult: the objects do not decrease significantly 
in size but begin to overlap 
and become increasingly deformed. While it seems that one can give to some extent ``physical"
values (by which it is meant: invariant under rescaling of the coupling) for the I(A) density and overlap 
if only fluctuations above 
a given wave-length are retained, 
these values depend on the latter and   
it appears difficult to select some ``more physical" sub-ensemble 
characterized by universal parameters.

The connection between topology and chiral properties is well observed at the level 
of the topological structure which define the topological sector: the instantons
which appear responsible for the net topological charge
 appear also strongly correlated with the Dirac
modes of small eigenvalue and large chirality. This connection can also be 
followed with the temperature.

The analysis of the MC configurations typically needs some procedure to
get rid of short range fluctuations. We have used here a ``scale controlled cooling algorithm", RIC, which has been developed to act as a gauge invariant 
low pass filter. RIC smoothes out fluctuations up to a chosen scale given as a
physical (dimensionfull) length and has no other parameter (like setting the
number of
cooling sweeps $n_c$ etc). If the smoothing scale is taken 
large enough this algorithm
reproduces standard cooling and it can be used under circumstances
to understand the effect of the latter in its dependence on  $n_c$ \cite{cpn}.
Using RIC  allows for an independent 
scaling check, especially in connection with
such quantities (like the I-A density) 
which appear to depend on smoothing, since
the smoothing scale is independent on the lattice spacing. In general
this smoothing procedure can be used to study the phenomena at a given 
physical scale and may represent therefore a good instrument in extracting
dynamical information from MC configurations.\par\bigskip

\no {\bf Acknowledgments:} Support from the DFG and 
from the NATO Advanced Research Workshop, Dubna 1999, is thankfully acknowledged.

\end{document}